\documentstyle[bo99,epsfig]{article}

\title{The broad band spectrum and variability of Seyfert 1 }
\author{Giorgio Matt}
\affil{Dipartimento di Fisica,
 Universit\`a degli Studi Roma Tre,
via della Vasca Navale 84, I--00146 Roma, Italy
 }

\begin{document}

\maketitle

\begin{abstract}
Recent results on the X--ray spectrum and variability of Seyfert 1 galaxies
are reviewed. New spectral
results from BeppoSAX observations are also presented and briefly discussed. 
\end{abstract}

\section{Introduction}

With {\it Chandra} just launched, and XMM and ASTRO-E due to follow suit 
soon, it is the right time to review what we have learned on Seyfert
1 galaxies from previous generation satellites like ASCA and BeppoSAX, 
and point out the still open questions, hopefully to be answered or at least
addressed by the new generation missions. In particular, I will concentrate
here on BeppoSAX results, not only because I worked myself on them but also
because it is only now that a general picture from these observations 
is emerging. Before discussing the
BeppoSAX observations, however, let me at least mention briefly the 
recent results from ASCA and XTE which I consider most significant.

\section{Recent results on Seyfert 1}

\subsection{Variability}

The most exciting result in recent years on variability in Seyfert Galaxies
is probably the discovery of periodicity in the flux of the Compton--thin
Seyfert 2 galaxy IRAS~18325-5925 (Iwasawa et al. 1998). The discovery has
been obtained with an imaging satellite, ASCA; when the field is
observed with a better
spatial resolution instrument (ROSAT), no evidence for a confusing source 
is found. The chance of the periodicity being due to 
a cataclysmic variable (like in the infamous case of NGC~6814, Madejski et al.
1993) or some other confusing source is therefore very low. As periodicity
has been observed so far only in one (or two, see below) sources,
it is possible that it is a 
transient event, e.g. a flaring region orbiting around the black hole
with a lifetime longer than usual (or being so strong to 
outshine all other flares simultaneously present). IRAS~18325-5925
has many characteristics typical of the subclass of Narrow Line Seyfert 1
Galaxies (see Boller, 1999, for a review): large amplitude 
variation, steep spectrum, possibly ionized disc. One can then argue that 
periodicities are better to be searched for in NLS1s rather than in
`classical' Seyferts. However, a recent claim of a 33~h periodicity in
MCG--6-30-15 by Lee et al. (2000) could indicate that periodicity may
be a not so rare phenomenon, after all, and that it has been generally 
missed so far due to lack of long enough observations. 

Thanks mostly to RXTE, long monitoring campaigns in X--rays simultaneously
with optical and UV have been performed on a few selected objects, with
quite surprising and important results. In NGC~7469 (Nandra et al. 1998),
NGC~5548 (Chiang et al. 1999) and NGC~3516 (Maoz et al. 2000), 
optical/UV and X--ray emissions seem
to be uncorrelated or, if any correlation exists, is the optical/UV that leads
X--rays, opposite to what expected in the reprocessing scenario.
This scenario (Collin--Souffrin 1991), 
in which optical and UV emissions arises as
reprocessing in the accretion disc of the hard X--rays, was suggested
to explain the too tight correlation (and lack of delays)
between optical and UV fluxes (Courvoisier \& Clavel 1991) in some Seyfert 1s 
(NGC~5548 being one of the best examples). So,
the situation is rather confused and puzzling, as in the same source
both the viscous dissipation and the reprocessing scenario appear to
be unviable.

Regarding even longer term variability, the 
most spectacular result recently has
been the switching off of the ``swan's song" galaxy, NGC~4051 (Guainazzi
et al. 1998). When observed by BeppoSAX in May 1998,
only the Compton reflection component and the iron line were visible,
while the primary radiation had disappeared. When put
in the context (Uttley et al. 1999), this observation demonstrated that the
observed reflected emission occurred in matter
distant at least a few light--months from the nucleus. So far, this is
the most convincing evidence for the presence of large amount of cold
circumnuclear matter in Seyfert 1 galaxies. 

\subsection{Warm absorbers}

Warm absorbers are detected in at least half of Seyfert 1s (Reynolds 1997, 
George et al. 1998), indicating that the covering factor of the ionized
material is fairly large. 
 The reason why warm absorbers are not observed in all
Seyferts is not clear. It may be due either to a spread in column density
and/or ionization structure,
or to a covering factor of the ionized matter less than unity. In the
first case, an improved sensitivity, as for instance that provided by XMM, is
necessary to search for small edges due either to small columns of oxygen or
to less abundant elements (e.g. Neon). In the second case, the ionized material
may be observed via emission lines (e.g. Netzer 1993), and high resolution
detectors (gratings, {\it Chandra} and XMM, or calorimeters, ASTRO-E) are
needed.

The most interesting recent discovery possibly related to
warm absorbers is the ~1 keV
absorption feature observed in some NLS1s 
(Vaughan et al., 1999a; Leighly 1999 and references therein). 
When interpreted as a blend of several resonant lines, mainly from 
iron L (Nicastro, Fiore \& Matt 1999), 
this implies that the ionization structure of narrow and broad lines 
Seyferts is rather different, not surprisingly given the different
ionizing continuum, especially in soft X--rays. Other possible explanations
include blueshifted absorption edges (Leighly et al. 1997) 
and oxygen smeared edge
from ionized discs (Fabian, private communication; see also Vaughan et al.
1999b). XMM should easily distinguish between competing models.

\subsection{The iron line}

The iron line in Seyfert 1 galaxies is usually broad (Nandra et al. 1997), 
likely originating in a relativistic disc (e.g. Fabian et al. 1995).
In the two best studied sources, 
MCG--6-30-15 (Tanaka et al. 1995; Guainazzi et al. 1999) and NGC~3516 (Nandra et
al. 1999), the observation of a double--horned and 
asymmetric line profile, expected in presence of fast orbital motion and strong
gravity effects (e.g. Fabian et al. 1989; Matt et al. 1992), makes this 
interpretation quite strong. 

In principle, from the line profile and short term variability 
in response to variations of the continuum (i.e. the so--called
reverberation mapping),
it is possible to determine the spin of the black hole and the geometry 
of the accretion disc and the illuminating source (e.g. Martocchia, Karas \&
Matt 1999 and references therein). 
Large sensitivity, coupled with good energy
resolution, it is required for these goals. It is possible that even the 
next missions will not sufficient for this purpose, and we have probably
to wait for Constellation--X (e.g. Young \& Reynolds 2000) 
and XEUS. It is worth noting, however, that already with ASCA it has been
possible to start studying variations of the iron line in MCG--6-30-15
(Iwasawa et al. 1996; 1999). Even if the results are not yet conclusive, 
these works demonstrated the potentiality of the method.
   
While most of the iron lines observed in Seyfert 1s can be explained in
terms of relativistic discs, it is not clear whether emission from more
distant matter (let us call it the ``torus") is also present. Evidence
for this component is scanty, and the only unambiguous cases so far are 
those of NGC~4151 (e.g. Piro et al. 1999)
and of NGC~4051 discussed in Sec.2.1. In the latter case, the discovery was 
due more to a stroke of luck rather than to systematic searches, and the 
probability to find other sources switched off seems not very high. More 
promising, in view of missions like XMM and expecially ASTRO--E with
large sensitivity coupled with good energy resolution, 
is to search for narrow iron
lines (as the torus should be at a distance such that relativistic effects
are negligible).

\subsection{The continuum}

While there is wide consensus on the fact that the power law index
(assuming that a power law is actually a good enough description of the
primary continuum: see Petrucci et al., 1999) is typically
around 1.8--2 (Nandra \& Pounds 1994) and that the Compton reflection
is a common ingredient, there are still several open questions: is there
a correlation between the photon index and the amount of Compton reflection,
as suggested by Zdziarski et al (1999)? How common is the soft excess
in Broad Lines Seyfert 1s (there is  actually no
doubt that it is common in
Narrow Lines Seyfert 1s)? What is the typical (if any) cut--off energy in 
the hard X--ray spectra? To address these questions, let me discuss
the BeppoSAX observations of a dozen of moderately bright Seyfert 1s.

\section{A BeppoSAX spectral survey of Seyfert 1}

In Table 1, the observed sources and the main spectral parameters are listed.
When the sources have been observed more than once, the results refer
to the combined spectrum if there is not spectral variability, or otherwise
to the observation when the source was brightest.

For all sources, the spectral model includes: a power law with exponential 
cut--off; a gaussian line (or a relativistic line, when it provides a 
significantly better fit); the Compton reflection continuum. A warm absorber
and/or a soft excess have been added if required by the data. The spectral
parameters reported in the Table refer to the best fit model, whatever its
complexity. 

\begin{table}
\caption{The best fit parameters for the Seyfert 1s observed by BeppoSAX. 
For the case of Mkn~841, see the caveats in the text.
All errors refer to 90\% confidence level for two interesting
parameters, i.e. $\Delta\chi^2$=4.6.}
\begin{center}
\scriptsize
\begin{tabular}{||c|c|c|c|c|c||}
\hline
\hline
~& ~& ~& ~ & ~ & ~\cr
Source & S.E. & $\Gamma$ & R & EW (eV) & E$_c$~(keV) \cr
~& ~& ~& ~ & ~ & ~\cr
\hline
~& ~& ~& ~ & ~ & ~\cr
NGC~5548$^{1,2}$  &      NO   & 1.63$^{+0.04}_{-0.03}$ &  0.44$^{+0.18}_{-0.18}$
& 120$\pm$40 &     160$^{+50}_{-70}$ \cr
~& ~& ~& ~ & ~ & ~\cr
\hline
~& ~& ~& ~ & ~ & ~\cr
NGC~3783$^1$  &    NO   & 1.71$^{+0.08}_{-0.08}$   & 0.37$^{+0.22}_{-0.22}$
& 190$\pm$65 &  160$^{+90}_{-60}$ \cr
~& ~& ~& ~ & ~ & ~\cr
\hline
~& ~& ~& ~ & ~ & ~\cr
NGC~7469$^1$  &    NO?   & 2.04$^{+0.05}_{-0.05}$  & 1.1$^{+0.4}_{-0.3}$
& 180$\pm$90 &  $>$~230 \cr
~& ~& ~& ~ & ~ & ~\cr
\hline
~& ~& ~& ~ & ~ & ~\cr
NGC~4151$^1$  &      ?    & 1.35$^{+0.20}_{-0.20}$  & 0.45$^{+0.20}_{-0.20}$
& 240$\pm$70 &  70$\pm$20  \cr
~& ~& ~& ~ & ~ & ~\cr
\hline
~& ~& ~& ~ & ~ & ~\cr
Fairall~9  &    ?      & 2.05$^{+0.11}_{-0.09}$ & 1.2$^{+0.7}_{-0.5}$
& 220$^{+60}_{-100}$ &  $>$~220 \cr
~& ~& ~& ~ & ~ & ~\cr
\hline
~& ~& ~& ~ & ~ & ~\cr
NGC~5506  &      ?    & 2.03$^{+0.09}_{-0.08}$  & 1.5$^{+1.2}_{-0.5}$
& 175$\pm$45 &   $>$~380 \cr
~& ~& ~& ~ & ~ & ~\cr
\hline
~& ~& ~& ~ & ~ & ~\cr
NGC~4593$^3$  &    NO     & 1.87$^{+0.05}_{-0.05}$ & 1.1$^{+0.4}_{-0.4}$
& 190$^{+90}_{-6}$ &  $>$~150 \cr
~& ~& ~& ~ & ~ & ~\cr
\hline
~& ~& ~& ~ & ~ & ~\cr
IC~4329A$^4$  &    NO      & 1.86$^{+0.03}_{-0.03}$  & 0.55$^{+0.15}_{-0.13}$
& 110$^{+50}_{-40}$ &  270$^{+170}_{-80}$ \cr
~& ~& ~& ~ & ~ & ~\cr
\hline
~& ~& ~& ~ & ~ & ~\cr
Mrk~509$^5$ &    Yes     & 1.58$^{+0.08}_{-0.09}$  &
0.6$^{+0.3}_{-0.3}$
& 110$^{+130}_{-60}$ & 70$^{+50}_{-20}$  \cr
~& ~& ~& ~ & ~ & ~\cr
\hline
~& ~& ~& ~ & ~ & ~\cr
MCG-8-11-11$^5$      &  NO       & 1.84$^{+0.05}_{-0.05}$ & 1.0$^{+0.6}_{-0.4}$+
& 130$^{+70}_{-50}$ &  170$^{+300}_{-80}$  \cr
~& ~& ~& ~ & ~ & ~\cr
\hline
~& ~& ~& ~ & ~ & ~\cr
MCG-6-30-15$^6$     &  NO       & 2.06$^{+0.03}_{-0.03}$ & 1.2$^{+0.4}_{-0.2}$
& 200$^{+50}_{-60}$ &  160$^{+130}_{-60}$  \cr
~& ~& ~& ~ & ~ & ~\cr
\hline
~& ~& ~& ~ & ~ & ~\cr
Mrk~841       & Yes?      & 2.16$^{+0.07}_{-0.05}$ & 3.9$^{+1.2}_{-1.2}$
& 290$^{+210}_{-220}$ &  $>$~150   \cr
~& ~& ~& ~ & ~ & ~\cr
\hline
~& ~& ~& ~ & ~ & ~\cr
NGC~3516$^{7}$ &    NO     & 2.04$^{+0.03}_{-0.03}$ & 1.4$^{+0.4}_{-0.4}$
& 100$^{+70}_{-60}$ &  $>$~350 \cr
~& ~& ~& ~ & ~ & ~\cr
\hline
\hline
\end{tabular}
\end{center}
$^1$ Piro et al. 1999; DeRosa et al. 1999; 
$^2$ Nicastro et al. 2000;
$^3$ Guainazzi et al. 1999a;
$^4$ Perola et al. 1999;
$^5$ Perola et al. 2000;
$^6$ Guainazzi et al. 1999b;
$^7$ Costantini et al. 2000.
\end{table}

In Fig.~1 the distributions of $\Gamma$, $R$ (the solid angle,
in units of 2$\pi$, 
subtended by the reflecting matter, assumed to be observed face--on) and
of the equivalent width of the iron line, are given. The results are
in agreement with those found by Nandra \& Pounds (1994) and
Nandra et al. (1997) and based on {\it Ginga} 
and ASCA observations, respectively.

\begin{figure}
\centerline{\epsfig{file=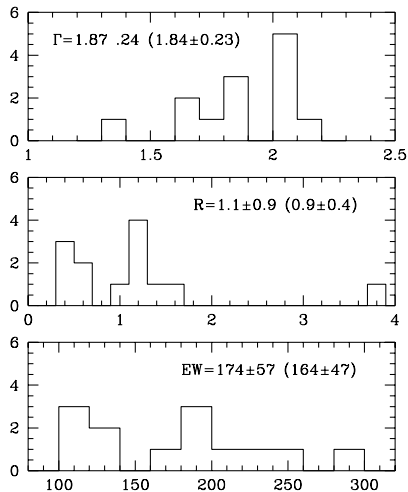, width=8cm}}
\caption[]{The distributions of power law indices, $\Gamma$, of the
amount of Compton reflection, $R$ and of the iron line EW. The average
values and standard deviations are labeled. In parentheses, the value 
excluding Mkn~841 are also given.}
\label{histogram}
\end{figure}

\subsection{The high energy cut--off}

Before the launch of BeppoSAX, a high energy cut--off was unambiguously 
detected only in one source, NGC~4151 (Jourdain et al., 1992).
From the analysis of the spectrum obtained summing several sources observed 
with OSSE, Gondek et al. suggested an average cut--off energy 
of a few hundreds keV. 

From Table~1, it can be seen that a high energy cut--off is positively
detected in about half of the sample.
The detected values range from 70 keV (NGC~4151) to 270 keV 
(IC4329A). When only a lower limit can be obtained,
the values are generally consistent with the measured ones.
The plot of the $e$-folding energies vs. the power law index is shown in 
Fig.~2. While there is no obvious correlation between the
two parameters, there is a tendency for flat sources to avoid high
values of the cut--off energy. Whether this reflects something intrinsic 
to the spectra, or simply the fact that it is easier to determine high energy
cut--offs in flat sources, cannot be said with the present data. 

\begin{figure}
\centerline{\epsfig{file=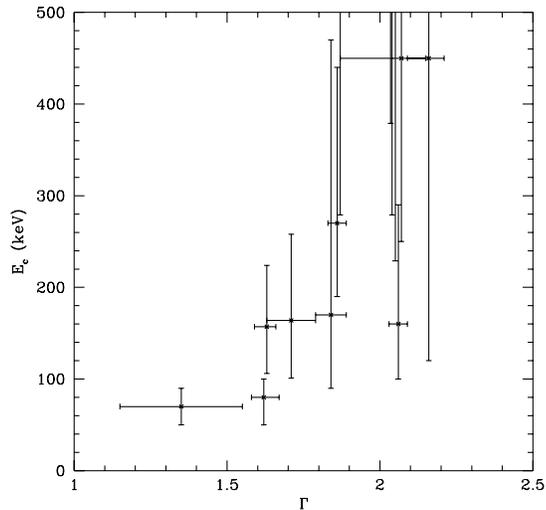, width=8cm}}
\caption[]{The high energy exponential cut--off, $E_C$, versus the photon
index $\Gamma$. }
\label{gammaec}
\end{figure}

\subsection{The soft excess}

Only two out of 13 sources (namely Mkn~509 and Mkn~841)
in the BeppoSAX sample discussed here present
evidence for a soft excess. Even if for three sources (NGC~5506 and NGC~4151,
which have a large intrinsic absorption, and Fairall~9, for which the LECS
instrument 
was switched off, for  technical problems, for  almost all the observation)
we cannot say anything, the soft excess appears to be the exception rather 
than the rule. 

For Mkn~509, the inclusion of a soft excess (in the form of a
steeper power law dominating below about 1 keV) improves significantly the
goodness of the fit and provides a more astrophysically
sound global solution (Perola et al. 2000).  
For Mkn~841, a warm absorber fits the spectrum 
almost as well as a black body soft excess, but from ASCA it seems
that the warm absorber is not very relevant in this object
(see e.g. Reynolds 1997). In both sources, the shape and 
luminosity of the soft component remains rather undetermined, and what
can be said is that a steepening of the spectrum occurs below
1 keV or so. Whether there is a curvature in the primary X--ray emission, or 
there are two different components, it is impossible to establish.

\subsection{The iron line and Compton Reflection}

In Fig.~3 the iron line EW is plotted against $R$. In the
reprocessing scenario, the two quantities are expected to be strongly
correlated (George \& Fabian 1991; Matt, Perola \& Piro 1991). 
The error bars, however, are too large to really check this
hypothesis. The relatively small probability (3.4\%) that they 
are not correlated is mainly due one point, i.e. Mkn~841 (but see
next paragraph). 

More interesting and instructive is the comparison between the average 
values of the two quantities (Fig.~1), as it can be used to estimate the iron
abundance, which turns out to be consistent (say, within a factor of 2 or so)
with the cosmic value.

\begin{figure}
\centerline{\epsfig{file=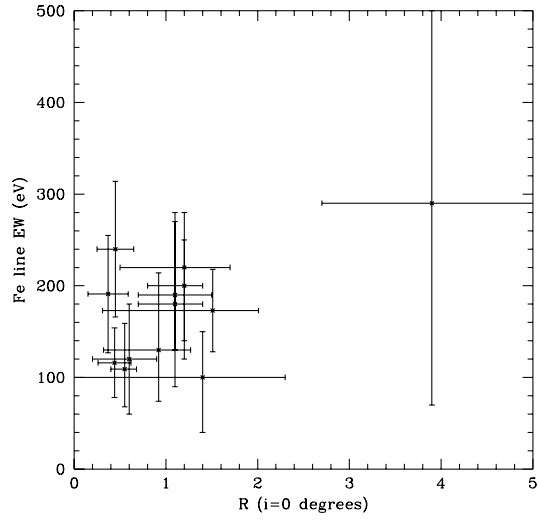, width=8cm}}
\caption[]{The iron line EW vs. $R$.}
\label{rew}
\end{figure}

\subsection{Are $\Gamma$ and $R$ correlated?}

In Fig.~4 $R$ is shown against the photon index. A correlation between
the two quantities is evident, the probability that they are not correlated
being only 1.3\% if Mkn~841 is excluded, and less than 0.1\% when it is
included. However, a word of caution is necessary here. The 
two parameters are strongly correlated in the fit procedure (see also 
the discussion in Nandra et al. 2000). If, for some
reason, the photon index is miscalculated, and e.g. a value steeper than real
is found, a larger $R$ is immediately obtained. Let me use the case of Mkn~841  
as an example. If the soft excess is modeled with a 
black body, the results for $\Gamma$, $R$ and the iron EW
listed in Table 1 and presented in the figures are obtained. An even steeper
spectrum ($\Gamma$=2.27) and, correspondingly,  larger $R$ (4.6) is obtained
with the warm absorber model. Conversely, 
if the soft excess is modeled by a power law, the photon index is significantly
flatter (($\Gamma$=1.75) and $R$ reduces to 1.5 (but the line EW
remains fairly large, e.e. 350 eV). 

Therefore, I believe that 
the issue of the reality of the $\Gamma$--$R$ correlation is still open, and
can be further addressed
only by much more sensitive missions like XMM (but the 
energy band may be not broad enough) or Constellation--X.

\begin{figure}
\centerline{\epsfig{file=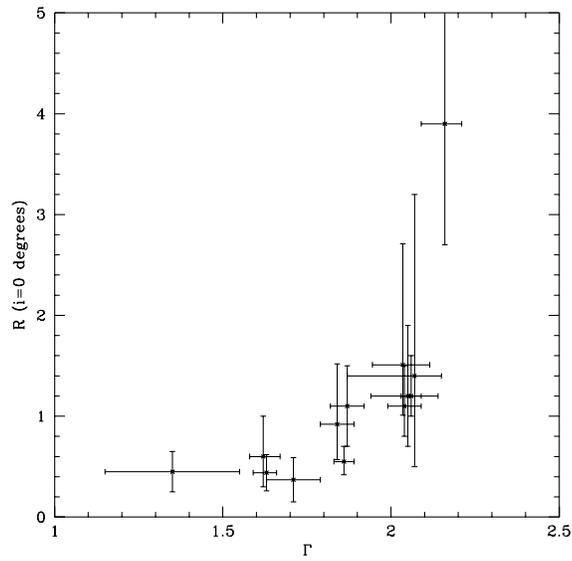, width=8cm}}
\caption[]{$R$ vs. $\Gamma$.}
\label{gammar}
\end{figure}


\section{Conclusions}

The main results from the BeppoSAX observations of a sample of Seyfert 1
Galaxies can be summarized as follows:

\begin{itemize}

\item{In most sources there is no evidence for a soft excess. In only
two sources, Mkn~509 and Mkn~841, a steepening of the spectrum below
about 1 keV seems required by the data. Whether it is due to a curvature
of the primary emission, or to an altogether distinct component, is
impossible to say.}

\item{ A high energy cut--off with $e$--folding energy typically in 
the range $\sim$100--200 keV
is observed, when the quality of the data is good enough.}

\item{The power law index, $\Gamma$, and the amount of reflection component
are clearly correlated, confirming the findings of Zdziarski et al. (1999).
However, it is possible that such a correlation is not real, 
but partly or entirely due to the fact that the two parameters are strongly
correlated in the fit procedure. Let us suppose that some 
values of $\Gamma$ are miscalculated, either because there are other 
components not taken into account, or because a power law is a too simple
model (e.g. Petrucci et al. 1999). In this case, a correlation between
$\Gamma$ and $R$ would be immediately found, as the quality of present (and of
course past) data is not sufficient to measure the amount of Compton
reflection independently of the shape of the primary continuum.}

\end{itemize}

\begin{acknowledgements}
The BeppoSAX results have been presented on behalf of large
teams led by G.C. Perola and L. Piro. I thank E. Costantini, A. De Rosa,
F. Nicastro, L. Piro and especially G.C. Perola for many useful discussions,
and for kindly providing the results of their analysis before publication.
I acknowledge financial support from 
the Italian Space Agency, and from MURST (grant {\sc cofin}98--02--32).
\end{acknowledgements}

 \end{document}